\documentclass[12pt]{article}

\usepackage{graphicx}
\usepackage{amsmath}
\usepackage{amsfonts}
\usepackage[mathscr]{euscript}
\usepackage{latexsym}

  \def\scr{\mathscr}

  \def\D{{\scr D}}
  \def\Db{\bar\D}

  \def\L{{\scr L}}
  
  \def\C{{\cal C}}

\newcommand{\N}{{\scr N}}

\def\pr{\partial}

\newcommand{\half}{{\textstyle \frac{1}{2}}}

\newcommand{\Leff}{\L_{\rm eff}}

\newcommand{\Geff}{\Gamma_{\rm eff}}

\newcommand{\dS}{\!\!{\rm d}^6z\,}
\newcommand{\dSb}{\!\!{\rm d}^6\bar z\,}
\newcommand{\dV}{\!\!{\rm d}^8z\,}

\newcommand{\al}{\alpha}
\def\da{{\dot\alpha}}
\def\be{\beta}
\def\db{{\dot\beta}}

\def\is{{^{\!(\sigma)}}}

\def\ga{\gamma}

\newcommand{\tfr}[2]{{\textstyle \frac{#1}{#2}}}
\newcommand{\fdq}[2]{\frac{\delta #1}{\delta #2}}
\def\ts{\textstyle}

\newcommand{\Liii}{\L_{3}}
\newcommand{\Liiib}{\L_{3}'}
\newcommand{\Liv}{\L_{4}}
\newcommand{\LRR}{\L_{R\bar R}}
\newcommand{\LW}{\L_{\rm Weyl}}
\newcommand{\Ltop}{\L_{\rm top}}

\newcommand{\Iiii}{I_{3}}
\newcommand{\Iiiib}{\bar I_{3}}
\newcommand{\IRR}{I_{R \bar R}}
\newcommand{\IW}{I_{\rm Weyl}}
\newcommand{\IWb}{\bar I_{\rm Weyl}}
\newcommand{\Itop}{I_{\rm top}}
\newcommand{\ws}{w^{(\sigma)}}
\newcommand{\wsb}{\bar w^{(\bar\sigma)}}

\newcommand{\utop}{u_{\rm top}}
\newcommand{\uiii}{u_3}
\newcommand{\uiiib}{u_3'}
\newcommand{\uW}{u_{\rm Weyl}}
\newcommand{\uRR}{u_{R\bar R}}

\newcommand{\rg}{{\rm geom}}

\newcommand{\W}{{\scr W}}

\begin{document}

\thispagestyle{empty}
\vspace*{-2cm}
\begin{flushright}
hep-th/9809090\\ NTZ 23/98\\ September 1998
\end{flushright}

\vspace{1cm}
\begin{center}
{\Large 
 \bf Geometrical Superconformal Anomalies}  

\vspace{1cm}

{\parindent0cm
Johanna Erdmenger\footnote{Supported by Deutsche Forschungsgemeinschaft,
e-mail: Johanna.Erdmenger@itp.uni-leipzig.de}
and 
Christian Rupp\footnote{Supported by Graduiertenkolleg
  "Quantenfeldtheorie: Mathematische Struktur und physikalische
Anwendungen",
e-mail: Christian.Rupp@itp.uni-leipzig.de}}

\vspace{0.7cm}

Institut f{\"u}r Theoretische Physik\\
Universit{\"a}t Leipzig\\
Augustusplatz 10/11\\
D - 04109 Leipzig\\
Germany
\end{center}

\vspace{2cm}

\centerline{\small \bf Abstract}\vspace*{-2mm}
{ \small 
We determine the structure of geometrical superconformal anomalies for
$N=1$ supersymme\-tric quantum field theories on curved superspace to
all orders in $\hbar$. For the massless Wess-Zumino model we show how
these anomalies contribute to the local Callan-Symanzik equation which
expresses the breakdown of superconformal symmetry in terms of the
usual $\beta$ and $\gamma$ functions.
}

\vspace*{30mm}
\begin{tabbing}
         \> Curved Superspace Background, Anomalies.
\end{tabbing}

\newpage

%%%%%%%%%%%%%%%%%%%%%%%%%%%%%%%%%%%%%%%%%%%%%%%%%%%%%%%%%%%%%%%%%%%%%%%

Superconformal symmetry transformations of a given $N=1$
supersymmetric quantum field theory are conveniently studied within
the superfield formalism by coupling the theory to a classical minimal
supergravity background.
For diffeomorphism inva\-riant models the superconformal transformations
are then determined by the super Weyl transformations.

We consider curved space backgrounds characterised by a Weyl invariant
supergravity prepotential $H_{\al\da}$, which couples to the
supercurrent of the theory, and a chiral compensator $\phi\equiv {\rm
  e}^J$ whose Weyl transformation is given by $\delta_\sigma J=\sigma$
with $\sigma$ chiral. 

For the four-dimensional massless Wess-Zumino model on curved superspace,
whose dynamical fields are denoted by $A$, $\bar A$ and the dynamical
coupling by $g$,  it was shown in
\cite{ERS} in a perturbative approach that the {\em local
  Callan-Symanzik equation}
\begin{align}
\left( w^\is -\gamma A \fdq{}{A} \right) \Gamma &= \left[ \beta^g
  \partial_g \Leff \right] \cdot \Gamma + C(H,J,\bar J)\,,
  \label{localCS} \\
w^\is &= \fdq{}{J} - A \fdq{}{A}\,, \label{wis}
\end{align}
holds to all orders in $\hbar$. Here $w^\is$ is the chiral Weyl
symmetry operator, $\Gamma$ the vertex functional and $\beta^g$ 
and $\gamma$ are the well-known
functions expressing the anomalies of scale invariance. The first term
on the right hand side is an insertion which is uniquely determined in terms of field
polynomials involving the dynamical fields. $C(H,J,\bar J)$ denotes
all purely geometrical superconformal anomalies.
The aim of this letter is to determine $C(H,J,\bar J)$ explicitly to
all orders in $\hbar$.\footnote{Our conventions are those of \cite{ERS}.} 

A basis for the geometrical terms $C(H,J,\bar J)$ in the local 
Callan-Symanzik equation (\ref{localCS}) is given by
\begin{equation}
\begin{array}{rcl}
\Ltop &=& \phi^3 \left( W^{\al\be\ga} W_{\al\be\ga} -
    (\Db^2+R)( G^a G_a - 2 R\bar R) \right) \\[0.2cm]
\LW &=& \phi^3 W^{\al\be\ga} W_{\al\be\ga} \\[0.2cm]
\LRR &=& \phi^3 (\Db^2+R) (R\bar R) \\[0.2cm]
\Liii &=&  \phi^3 R^3 \\[0.2cm]
\Liiib &=& \phi^3 (\Db^2+R)\bar R^2  \\[0.2cm]
\Liv &=& \phi^3 (\Db^2+R) \D^2 R \equiv \phi^3 \Box R  \, .
\end{array} \label{geombasis}
\end{equation}
$W_{\al \be \gamma}$, $G_a$ and $R$ are the supersymmetric generalisations
of the Weyl tensor, Ricci tensor and Ricci scalar respectively.
$G_a$ is a real vector superfield, while $W_{\al \be \gamma}$ and $R$
are chiral with their antichiral partners given by
$\bar W_{\da \db \dot{\gamma}}$, $\bar R$. 
These structures are
discussed further in textbooks like \cite{bk}.
The supercovariant derivatives are denoted by $\D$, $\bar \D$; $(\bar \D^2+R)$ is
the chiral projection operator.
For the integrals of the basis  (\ref{geombasis}) we define
\begin{equation}
\begin{array}{rclcrcl}
\int\dS\Ltop &=& \Itop & \qquad&\int\dS \LW &=& \IW \\[0.2cm]
\int\dS\LRR &=& \IRR & & \int\dS\Liv &=& I_4 \\[0.2cm]
\int\dS\Liii &=& \Iiii & & \int\dS\Liiib &=& \bar\Iiii\,,
\end{array}
\end{equation}
such that we obtain
\begin{equation}
\begin{array}{rcl}
\Itop &=& \int\dS \phi^3 W^{\al\be\ga} W_{\al\be\ga} - \int\dV E^{-1}
( G^a G_a - 2R\bar R ) \\[0.2cm]
\IW &=& \int\dS \phi^3 W^{\al\be\ga}W_{\al\be\ga}\\[0.2cm]
\IRR &=& \int\dV E^{-1} R\bar R \\[0.2cm]
\Iiii &=& \int\dS \phi^3 R^3 \\[0.2cm]
\Iiiib &=& \int \dSb  \bar \phi^3 \bar R^3 \\[0.2cm]
I_{4} &=& 0 \, ,
\end{array} \label{geomint}
\end{equation}
with $\int\dV E^{-1}$ the integration measure for curved superspace.
$\Ltop$ and $\LW$ are the supersymmetric Gau\ss-Bonnet and Weyl densities,
such that in an expansion around flat space we have
\begin{subequations}
\begin{gather} \label{topzero}
\IW - \IWb = \Itop - \bar\Itop = 0 \, , \\ \label{topzerob}
\Itop + \bar \Itop = 0 \, .
\end{gather} 
\end{subequations}
The basis (\ref{geombasis}) has been investigated using 
algebraic renormalisation in \cite{bonora} to first order in $\hbar$.
In \cite{bonora} it was found that only the Gau\ss-Bonnet and Weyl
terms are anomalies,
while $\Liii$, $\Liiib$ and $\Liv$ may be removed by suitable redefinitions
of the vertex functional, and $\LRR$ is eliminated by Wess-Zumino consistency.
Here we extend this analysis to all orders in $\hbar$ and discuss the
consequences for the local Callan-Symanzik equation (\ref{localCS}).

For the Weyl transformation properties of the geometrical terms
we have
\begin{equation}
\begin{array}{lclclcl}
\ws \Iiii &=& -3\Liii & \qquad & \wsb \Iiii &=& \phantom{-}3\bar \Liiib \\[0.2cm]
\ws \bar \Iiii &=&  \phantom{-}3\Liiib & & \wsb \bar\Iiii &=& -3\bar \Liii \\[0.2cm]
\ws \IRR &=&  \phantom{-3}\Liv & & \wsb \IRR &=&  \phantom{-3}\bar\Liv \\[0.2cm]
\ws\IW &=& 0 & & \wsb \IW &=& 0\\[0.2cm]
\ws\Itop &=& 0 & & \wsb \Itop &=& 0 \, .
\end{array} \label{weyltranstab}
\end{equation}
By inserting the Weyl transformation parameters $\sigma$, $\bar \sigma$
appropriate
for $R$ transformations and dilatations \cite{ERS} into
the Weyl operator 
\[\W = -i \int\dS \sigma w^\is -i \int\dSb \bar\sigma \bar
w^{(\bar\sigma)}\,,\]
we obtain global Ward operators
\begin{align}
\W^R &= \tfr{2}{3} \int\dS w^\is - \tfr{2}{3} \int\dSb \bar
w^{(\bar\sigma)} \,, \label{Rdef}\\
\W^D &= i \int\dS w^\is +i  \int\dSb \bar
w^{(\bar\sigma)} \label{Ddef}
\end{align}
for global R transformations and dilatations respectively.
By applying these operators to
the integrated expressions (\ref{geomint}), we find that all these
expressions are invariant under $R$ transformations and
dilatations except $\Iiii$ and $\Iiiib$, for which
\begin{gather}
\W^R (\Iiii + \Iiiib) = -4 (\Iiii - \Iiiib) \, , \qquad \W^D \Iiii =0
\, , \qquad \W^D \Iiiib = 0 \, .
\end{gather}

Within the BPHZ approach pursued here \cite{BPHZ}, 
the renormalisation process necessitates the introduction of an
auxiliary mass term \mbox{$M(s-1) \left(\int\dS \phi^3A^2 + \int\dSb
  \bar\phi^3 \bar A^2\right)$} even within the massless theory
\cite{quantum}. 
This term is at the origin of potential anomalies since it leads to
symmetry breaking hard insertions involving
\begin{equation}
\left[ M(s-1) \phi^3 A^2 \right] \cdot \Gamma \,.\label{massterm}
\end{equation}
Additional information is needed to evaluate such terms before the
limit $s=1$ may be taken since the parameter $s$ participates in the
subtraction process similarly to an external momentum. This information is
provided by the Zimmermann identities \cite{zi} according to which terms
like (\ref{massterm}) may be expressed by a basis of local field
polynomial insertions. In our case this leads to
\begin{equation}
\left. \left[ M(s-1) \phi^3 A^2 \right] \cdot \Gamma\right|_{s=1} =
[Z_{\rm dyn}] \cdot \Gamma + Z_{\rm geom}\,, \label{ZI}
\end{equation}
where the terms $Z_{\rm dyn}$ involving dynamical fields have been
discussed in \cite{ERS} and the geometrical contributions 
$Z_{\rm geom}$ are given by 
\begin{equation}
Z_{\rm geom} = \utop \Ltop + \uW \LW + \uRR \LRR + \uiii \Liii +
\uiiib \Liiib + u_4 \Liv  \label{Zgeom}
\end{equation}
in terms of the basis (\ref{geombasis}).
The coefficients $u$ are power series in $\hbar$.

In the BPHZ approach, vertex functions are calculated from the {\em
  effective action} $\Geff$ in agreement with the forest formula.
Here we have
\begin{equation}
\Geff = \Geff^{\rm dyn} + \Geff^\rg \,. \label{Geff}
\end{equation}
We specify the geometrical contribution $\Geff^\rg$ in such a way that
the removable symmetry breaking terms arising from the Zimmermann
identity may be cancelled,
\begin{equation}
\Geff^\rg = \tfr{1}{8} \lambda_3 (I_3+\bar I_3) + \tfr{1}{8}
\lambda_{R\bar R} I_{R \bar R} \,.
\end{equation}
With these results we obtain for the Weyl transformation of the
Wess-Zumino vertex functional $\Gamma$ on curved superspace using
(\ref{weyltranstab}) and (\ref{ZI})
\begin{equation}
\left. w^\is \Gamma \right|_{s=1} = -\tfr{3}{2} \left. 
[S_{\rm dyn}]\cdot \Gamma\right|_{s=1} -\tfr{3}{2} S_{\rm geom}
\label{weylward} 
\end{equation}
with $S_{\rm dyn}$ as determined in \cite{ERS} and\footnote{To first order in
  $\hbar$ the coefficients $u$ in (\ref{Sgeom}) have been calculated explicitly
  in a different approach \cite{bma}.}
\begin{align}
S_{\rm geom} &= \tfr{1}{12} \Big(\, \utop\Ltop + \uW\LW + \uRR \LRR  
\nonumber \\ 
&\quad\quad\quad 
+(\uiii+3\lambda_3) \Liii +  (\uiiib -3\lambda_3) \Liiib 
+  (u_4-\lambda_{R\bar R})\Liv \,\Big) \,. \label{Sgeom}
\end{align}

Due to (\ref{topzero}) the
only terms in (\ref{weylward}) contributing to global $R$
transformations (\ref{Rdef}) are
\begin{gather}
\W^R \Gamma \Big|_{s=1} = -{\ts \frac{1}{12}}  (u_3 - u'_3 + 6 \lambda_3) 
(\Iiii - \Iiiib )  \, ,
\end{gather}
such that when setting
\begin{gather} \label{Rinv}
\lambda_3 = {\ts \frac{1}{6}}  (u'_3 - u_3)
\end{gather}
we have global $R$ invariance. At the local level however, $R$ symmetry
is anomalous since the non-integrated terms (\ref{Sgeom}) and their
antichiral partners contribute to the divergence of the supercurrent.

The global Callan-Symanzik equation obtained by applying the
operator $\C \equiv m\partial_m + \beta^g \partial_g -\gamma \N$ to
$\Gamma$, with $\N=\int\dS A\fdq{}{A}+\int\dSb \bar A \fdq{}{\bar A}$
the operator counting matter legs, reads 
\begin{gather}
\C \Gamma \Big|_{s=1, H=J=0} = 0
\end{gather}
on flat space.
This equation acquires anomalies on curved superspace as well, 
which are of the form
\begin{align} 
\C \Gamma \Big|_{s=1} & = \Delta^{\rg}_{\C} \, , \label{CS} \\[0.2cm]
\Delta^{\rg}_{\C} & = -{\ts \frac{1}{8}} (1-2\gamma) \uW(\IW + \IWb) + {\ts \frac{1}{8}}
( \beta^g \pr_g \lambda_{R\bar R} - 2 (1-2\gamma)\uRR )  \IRR 
\nonumber\\& \phantom{+} + 
{\ts \frac{1}{8}} (\beta^g \pr_g \lambda_3 - (1-2\gamma)(\uiii +\uiiib) )
(\Iiii + \Iiiib) \, . \label{CSgeom}
\end{align} 
Note that due to (\ref{topzerob}) the Gau\ss-Bonnet term does not contribute
to the anomalies of the global Callan-Symanzik equation  in
agreement with the topological nature of this term  from which follows 
its scale independence. 
Furthermore, using the consistency condition
$[\W^R,\C]\Gamma=0$ we may show that the coefficient of the $\Iiii + \Iiiib$
term in (\ref{CSgeom}) vanishes,
\begin{equation} \label{iiicond}
\beta^g \pr_g \lambda_3 - (1-2\gamma)(u_3 +u'_3) = 0 \, ,
\end{equation}
such that (\ref{CSgeom}) reduces to
\begin{gather}
\Delta_{\C}^\rg = -{\ts \frac{1}{8}} (1-2\gamma)\uW (\IW + \IWb) +
{\ts \frac{1}{8}} ( \beta^g \pr_g \lambda_{R\bar R} - 2 (1-2\gamma)
\uRR )  \IRR  \,.
\end{gather}

Similarly we determine  the
geometrical terms $C(H,J,\bar J)$ in the local Callan-Symanzik equation
(\ref{localCS}). For the R invariant theory (\ref{Rinv}), they are given by 
\begin{align} 
\left(w\is-\gamma A\fdq{}{A}\right)\Gamma \Big|_{s=1} 
&= - \beta^g [\partial_g\Leff]\cdot\Gamma \Big|_{s=1} + C(H,J,\bar J) \,,
 \label{localCSgeom} \\[0.2cm]
C(H,J,\bar J) &=   
-\tfr{1}{8} (1-2\gamma) \utop
\Ltop - \tfr{1}{8} (1-2\gamma) \uW \LW \nonumber\\
&\quad + \tfr{1}{8} \left( \half \beta^g\partial_g \lambda_{R\bar R} -
  (1-2\gamma) \uRR \right) \LRR \nonumber \\
& \quad +\tfr{1}{8} \left( \half(\uiii+\uiiib) -2\gamma \uiiib \right)
(\Liii-\Liiib) 
\, , \end{align}
with the geometrical part of $\Leff = \L_{\rm dyn}+\L_\rg$ given by 
\begin{equation} \label{Leffbar}
\L_{\rm geom} \equiv { \ts \frac{1}{8}} \lambda_3 \Liii + { \ts \frac{1}{16}}
\lambda_{R\bar R} \LRR \, .
\end{equation}
Here we have chosen $\lambda_{R\bar R} = (1-2\gamma)u_4$ in order to
absorb the breaking term involving $\Liv$.
                   
Integrating (\ref{localCSgeom}) over chiral superspace and adding the
complex conjugate we recover the global Callan-Symanzik equation
(\ref{CS}) by virtue of (\ref{Ddef}) and of $\W^D\Gamma = im\partial_m
\Gamma$. 
The coefficients of the superconformal anomalies given by $C(H,J,\bar
J)$ are in general non-zero to any given order in $\hbar$. However
there are simplifications in special cases.
To first order in $\hbar$ the consistency equation
\begin{equation}
[w^\is(z), \bar w^{(\bar \sigma)}(z')] \Gamma =0 \label{consistency}
\end{equation}
implies
\begin{equation}
\uiii+\uiiib = O(\hbar^2), \qquad \uRR = O(\hbar^2)\,,
\label{consistency2}
\end{equation}
such that we obtain
\begin{equation}
\left(w\is-\gamma A\fdq{}{A}\right)\Gamma \Big|_{s=1} 
= - \beta^g [\partial_g\Leff]\cdot\Gamma \Big|_{s=1} -\tfr{1}{8} \utop
\Ltop -\tfr{1}{8} \uW \LW +O(\hbar^2)\,.
\end{equation}
Furthermore at fixed points where $\beta^g=0$ it is straightforward to
determine the consequences of the consistency condition
(\ref{consistency}) to all orders in $\hbar$ since no double
insertions appear when applying $w^\is$ to the r.h.s. of
(\ref{localCSgeom}) in this case. Therefore for $\beta^g=0$ the
relations (\ref{consistency2}) are valid to all orders in
$\hbar$, and (\ref{localCSgeom}) simplifies to 
\begin{align} 
\left(w\is-\gamma A\fdq{}{A}\right)\Gamma \Big|_{s=1} 
&=
-\tfr{1}{8} (1-2\gamma) \left( \utop
\Ltop + \uW \LW \right) \nonumber\\
&\quad -\tfr{1}{4} \gamma \uiiib (\Liii-\Liiib)\,, 
\end{align}
with $\gamma\uiiib=O(\hbar^2)$.
For the trivial fixed point of the Wess-Zumino model where
$g=\gamma=0$ this reduces to
\begin{equation}
w^\is \Gamma|_{s=1} = -\tfr{1}{8} \utop\Ltop -\tfr{1}{8}
\uW\LW\,. \label{trivial} 
\end{equation}
For the Wess-Zumino model this confirms that at the fixed point, the
structure of correlation functions involving the supercurrent is
entirely determined by (\ref{trivial}), as was recently discussed in
\cite{Osborn}.

\end{document}